\documentclass[%
 reprint,
showpacs,preprintnumbers,
 amsmath,amssymb,
 aps,
prbe,
]{revtex4-1}

\usepackage{graphicx}
\usepackage{dcolumn}
\usepackage{bm}

\begin{document}

\preprint{APS/123-QED}

\title{An exact alternative solution method of 1D Ising model with Block-spin transformation at $H=0$}

\author{Tuncer Kaya}%
\email{tkaya@yildiz.edu.tr}%
\affiliation{Physics Department, Y$\i$ld$\i$z Technical University, 34220 Davutpa\c sa-Istanbul/Turkey\\
              }

\date{\today}

\begin{abstract}
An alternative exact explicit solution of 1D Ising chain is
presented without using any boundary conditions (or free boundary
condition) by the help of applying successively block-spin
transformation. Exact relation are obtained between spin-spin
correlation functions in the absence of external field. To
evaluate average magnetization (or the order parameter), it is
assumed that the average magnetization can be related to
infinitely apart two spin correlation function as
$<\sigma>^{2}=<\sigma_{0}\sigma_{N}>$. A discussion to justify
this consideration is given in the introduction with a relevant
manner. It is obtained that $<\sigma_{0}\sigma_{1}>=\tanh{K}$,
which is exactly the same relation as the previously derived
relation by considering the configurational space equivalence of
$\{\sigma_{i}\sigma_{i+1}\}=\{s_{i}\}$ and the result of transfer
matrix method in the absence of external field. By applying
further block-spin transformation, it is obtained that
$\{\sigma_{0}\sigma_{j}\}=(\tanh{K})^{j}$, here $j$ assumes the
values of $j=2^n$, here $n$ is an integer numbers. We believe that
this result is really important in that it is the only exact
unique treatment of the 1D Ising chain beside with the transfer
matrix method. It is also pointed out the irrelevances of some of
the alternative derivation appearing in graduate level text books.
The obtained unique correlation relation in this this study leads
in the limit of $N\rightarrow\infty$ to
$<\sigma_{i}>=(\tanh{K})^{N/2}$, indicating that the second order
phase transition is only possible in the limit of
$K\rightarrow\infty$. The obtained results in this work are
exactly the same as those of obtained by the transfer matrix
method which considers periodic boundary condition. Therefore, it
may be also claimed that the boundary conditions, either periodic
or free, do not affect the phase transition character of 1D Ising
chain in the absence of external field.

\pacs{75.10.Hk, 05.50.+q} keywords: {Ising model, Block-spin
transformation, Critical phenomena}
\end{abstract}
\maketitle


\section{\label{sec:level1}INTRODUCTION}
The Ising model \cite{Ising,huang} is a well-known and
well-studied model of magnetism. Its seemingly simplicity has been
attracted the concerted attention of physicists to investigate
mostly the second order phase transition phenomena for almost a
century. Beside, it also has played a key role in describing some
complex systems such as binary allays, long range interactions
\cite{cannas}, strong and long range correlations, random
disordered ferromagnet \cite{carlos}, neural networks, molecular
biology and also even in investigation of the complex behavior of
financial and economic markets \cite{stauffer}. As it is a
century-old problem and a large number of theories have been
developed and discussed, it is difficult to give an over view of
all the method in a research paper, we like, however, mention a
few important developments and methods in the following
paragraphs.

The 1D Ising model is the simplest case of the Ising model and its
treatment, therefore,  include the less mathematical
complications. It is just a chain of $N$ spin, each spin
interacting only with its two nearest-neighbors and with an
external magnetic field. Studying 1D Ising model is important in
that it is a good example of interacting 1D model system which can
be solved exactly. Indeed, Ising by himself solved it by a
combinatorial approach \cite{Ising,seht} and found that there was
no phase transition at a finite temperature. Later, the
combinatorial approach has been also applied to higher dimensional
Ising systems \cite{kac,potts,baker}. The mathematical
complication created by its application of higher Dimensional
Ising system naturally leaded to look for other simpler
approaches.

One such approach used in the treatment of Ising systems is the
well known transfer matrix method which was developed mainly by
Kramers and Onsager \cite{kramers,onsager}. The application of
this approach to 1D Ising model is quite easy while its
application to the 2D model, even in the absence of external
magnetic field, turned out to be an extremely difficult task.
However, Onsager achieved to obtain an explicit expression for the
free energy in zero field and thereby established the nature of
specific-heat singularity in 2D. The singularity in heat-capacity
interpreted as the indication of a phase transition at a finite
temperature. Apparently, the absence of external field simplify
the mathematics, but makes it impossible to obtain the order
parameter to investigate phase transition properly. Therefore, the
absence of external field creates so much conceptual complication
in the treatment of Ising system. This complication might be the
implicit motivation of Mc Coy's works of cumbersome calculation of
2D correlation functions \cite{McCoy1,McCoy2} explicitly. Indeed,
how to define properly  the order parameter in the absence of the
external magnetic field \cite{Kaufman1,Kaufman2}is going to be
one of the main motivation of this paper.

Other then these two main approaches, namely combinatorial and
transfer matrix, there are also different approaches such as
star-triangle transformation \cite{baxter1}and real space
renormalization group \cite{Kadanof,Goldenfeld,Ghosh1,Ghosh2}.
Their main concerns, however, are to obtain the critical
parameters of the systems instead of determining explicit analytic
function of order parameter.

As far as 1D Ising model is concerned, the simplest method to
obtain the nearest neighbor correlation function can be the use of
the configurational space equivalence of the interaction term with
a more simpler equivalent configuration. But its application to
calculate the correlation functions other then the nearest
neighbor correlation function is both irrelevant and wrong. This
wrong treatment appears even in the graduate level text book
\cite{beale,Goldenfeld}. We are going to, of course, give clear
explanation of this point in the following section. But, let me
just say now that their only aim is to get the already known
correlation function expressions obtained by the transfer matrix
method at zero field.

As point out in the above, a significant amount of different
approaches have been developed and worked out. So what is the
significance of introducing a new one. Of course, this is a good
question and must to be responded from the outset. Of course,
there might be a lot of different responds to this question, but
let me mention a few of them. First there is no exact treatment of
1D Ising model other than the transfer matrix method. Second,
there is no exact treatment of 3D Ising model since applications
of all the approaches produce non-followable mathematical
complications. Third, its application to higher dimensional Ising
model might create less mathematical complications. Forth, it does
not involve transfer matrices or combinatorial arguments. Fifth,
it does not create a new physical assumptions. Sixth, it can be
solved under by free boundary (or open boundary) condition, thus
it does not create conceptual confusion in the interpretation of
phase transition as in the case of periodic boundary condition.
Sixth, it allows us to deeply understand the physics lying behind
the 1D Ising model. Last but not least, it is always nice to
revisit an old problem from scratch and from different aspects.

Due to the above significant motivation and relevant hope, we are
going to present a new approach to obtain analytically zero field
correlation functions of The 1D Ising model in this paper. To
derive explicit expressions of the correlation functions, we are
going to exploit mainly the successive applications of Block-spin
transformations. It is important to mention that the Block-spin
transformation is also the starting point of real space
renormalization group theory and plays really important role to
obtain the critical values. However, the main concern of the
present paper is to obtain analytically the correlation functions
of 1D Ising model in the absence of external magnetic field by
repeated use of Block-spin transformation. Therefore,  it might be
proper to name this approach as Block-spin transformation method.

As it is going to be seen in the following section, the new
approach might be considered mathematically less complicated than
both of the transfer matrix and the combinatorial methods.
Although, we did not study the Ising system except for 1D case in
the present paper, we believe intuitively that its applications to
higher dimensional Ising system might be a lot less complicated
also. In addition, its application to 3D Ising model model might
be possible, at least with a proper approximate manner. Therefore,
we consider the new approach presented in the present paper is
relevant and promising.

This paper organized as follows. In the next section, the main
objective is going to be to derive analytical expressions of
correlation functions. In the same section, the irrelevance of the
derivation of correlation functions based on configurational space
equivalence appearing in the cited text books is going to be
explained as rigourously as possible. The last section is going to
be devoted to the discussion of a proper definition of order
parameter at zero magnetic field. Some concluding remarks are also
going to be given in the same section.

\section{\label{sec:level1}Obtaining Exact correlation function for
1D Ising model with Block-spin transformation}

The solution of 1D chain with transfer matrix method predicts that
the correlation function between two spins equals
$<\negthinspace\negthinspace\sigma_{i}\sigma_{i+j}\negthinspace\negthinspace>=\tanh(K)^J$
as the external magnetic field goes to zero. It is important to
point out that in the original solution of the 1D chain, Ising
preferred to apply periodic condition. There is also an exact
treatment of nearest neighbor correlation function for 1D Ising
chain based on the configurational space equivalence of
$\{\sigma_{i}\sigma_{i+1}\}=\{s_{i}\}$, here each $s_{i}$ assumes
$\pm1$ values. This configurational space symmetry leads easily to
the nearest neighbor correlation relation,
$<\negthinspace\sigma_{i}\sigma_{i+1}\negthinspace>=\tanh(K)$,
with free boundary condition. This result is very important in
that it is the only exact result obtained without using the
transfer matrix method.

As pointed out in the introduction section, the main purpose of
this work is to find exact correlation function relations for 1D
Ising model without using the transfer matrix method. This is,
however, by no means immediately obvious and an examination of the
literature shows that there is no alternative exact method for the
solution of 1D Ising model except for some irrelevant and
unacceptable treatments appearing in some graduate level text
\cite{Goldenfeld} books.

In these books, they consider random coupling between nearest
neighbor spins interaction and obtain the following correlation
function relations
\begin{eqnarray}
<\sigma_{i}\sigma_{i+k}\negthinspace\negthinspace>=\negthinspace<\negthinspace\sigma_{i}\sigma_{i+1}\negthinspace\negthinspace>
<\negthinspace\sigma_{i+1}\sigma_{i+2}\negthinspace>\negthinspace....
\negthinspace<\negthinspace\negthinspace\sigma_{k+1}\sigma_{i+k}\negthinspace\negthinspace\negthinspace>
\end{eqnarray}
This equation is, of course, true since sice all the coupling
constants are different. Using the notation
$<\sigma_{i}\sigma_{i+n}>=C(K_{n})$, the last equation can also be
written as
\begin{eqnarray}
<\sigma_{i}\sigma_{i+k}>=C(K_{i})C(K_{i+1})....C(K_{i+k-1})C(K_{i+k})
\end{eqnarray}
Then assuming all the random coupling are the same $K_{i}=K$, they
obtain $<\sigma_{i}\sigma_{i+k}>=C^{k}(K)$. Since $C(K)=\tanh(K)$,
they obtain the aimed relation
\;\;\;\;\;\;\;\;\;\;\;\;\;\;\;\;$<\negthinspace\sigma_{i}\sigma_{i+k}\negthinspace>=\tanh^{k}(K)$,
which is equivalent to the result obtained from transfer matrix
method. Although, the results of both method are the same, there
is very irrelevant and unacceptable step in the later treatment.
The irrelevancy can be noticed quite easily if we consider the
equal the coupling constants case. In this case it is impossible
to obtain the relations as presented by the Eq.(1) and Eq.(2).
Therefore, the assumption of the equal random coupling constants
in their derivations are irrelevant and wrong.  One can easily see
the irrelevancy by considering the applications of Eq.(1) to other
Ising system such as 1D Ising chain in the presence of external
magnetic field and 2D Ising system. So Eq.(1) is wrong for equal
coupling strengths and the obtained correlation relations by its
application are just mathematical hallucinations. Therefore, it
might be still important to obtain the correlation functions of 1D
Ising chain with a unique method other than using the transfer
matrix method.

In an effort to find an alternative exact solutions for 1D Ising
model in the absence of external field, we think that using the
well known Block-spin transformation of the original lattice is
quite relevant as it is particularly suitable for such a purpose.
Thus, let us write the 1D Ising chain partition function in the
following form,
\begin{eqnarray}
Q=\sum_{\{\sigma,\sigma^{'}\}}e^{K\sum\sigma_{i}^{'}(\sigma_{i-1}+\sigma_{i+1})}.
\end{eqnarray}
This partition function can be also expressed equivalently as
\begin{eqnarray}
Q=\sum_{\{\sigma,\sigma^{'}\}}\prod{e^{K\sigma_{i}^{'}(\sigma_{i-1}+\sigma_{i+1})}}
\end{eqnarray}
If two values of $\sigma=\pm1$ are inserted in the product, the
partition function becomes
\begin{eqnarray}
Q=\sum_{\{\sigma\}}\prod{[e^{K(\sigma_{i-1}+\sigma_{i+1})}+e^{-K(\sigma_{i-1}+\sigma_{i+1})}]}
\end{eqnarray}
the exponential terms in the product can be expressed as
$2\cosh(K(\sigma_{i-1}+\Sigma_{i+1}))$, thus the partition
function turns out to be
\begin{eqnarray}
Q=\sum_{\{\sigma\}}e^{\sum{\ln[2\cosh
K(\sigma_{i-1}+\sigma_{i+1})]}}
\end{eqnarray}
Expending the term,$\ln2\cosh(K(\sigma_{i}+\Sigma_{i+1}))$, into
series, it easy to see that the final partition function can also
be expressed as
\begin{eqnarray}
Q=\sum_{\{\sigma\}}\prod
f_{1}(K)e^{g_{1}(K)\sigma_{i-1}\sigma_{i+1}}
\end{eqnarray}
The function can be determined either by direct series expansion
or comparing the last equation with Eq.(7). We find that the
former is cumbersome while the latter is quite easy. Thus the
comparison produces the following equations for $f_{1}(K)$ and
$g_{1}(K)$,
\begin{eqnarray*}
&&f_{1}(K)=2[\cosh(2K)]^{1/2},{}\nonumber
\\&&
g_{1}(K)=\frac{1}{2}\ln[\cosh(2K)].
\end{eqnarray*}
With these relations, the Eq.(3) turns out to be
\begin{eqnarray}
Q=\negthinspace
\sum_{\{\sigma\}}f_{1}(K)^{N/2}e^{g_{1}(K)\sum\sigma_{i-1}\sigma_{i+1}}.
\end{eqnarray}
Taking the natural logarithm and calculating the derivatives of
the both side of the equation with respect to $K$ leads to
\begin{eqnarray*}
<\sigma_{i}^{'}\sigma_{i-1}>=\frac{1}{2}\frac{1}{f_{1}(K)}\frac{d
f_{1}(K)}{dK}+\frac{1}{2}\frac{d
g_{1}(K)}{dK}<\sigma_{i-1}\sigma_{i+1}>
\end{eqnarray*}
where $<\sigma_{i}^{'} \sigma_{i+1}>$ is the nearest neighbors
correlation function while $<\sigma_{i-1}\sigma_{i+1}>$ is the
next nearest neighbors correlation function. Substituting the
derivative in the above equation leads to
\begin{eqnarray}
<\sigma_{i}^{'}\sigma_{i}>=\frac{1}{2}\tanh(2K)+\frac{1}{2}\tanh(2K)<\sigma_{i}\sigma_{i+1}>
\end{eqnarray}
This equation can also be written as
\begin{eqnarray}
\frac{<\sigma_{i}^{'}\sigma_{i+1}>}{1+<\sigma_{i-1}\sigma_{i+1}>}=\frac{1}{2}\tanh(2K)
\end{eqnarray}
It is not hard to see that the function $\frac{1}{2}\tanh(2K)$ is
equal to $\frac{\tanh(K)}{1+\tanh^{2}(K)}$. Substituting the
equivalent form in the above equation, it can be written as
\begin{eqnarray}
\frac{<\sigma_{i}\sigma_{i+1}>}{1+<\sigma_{i-1}\sigma_{i+1}>}=\frac{\tanh(\small{K})}{1+\tanh^{2}(\small{K})}
\end{eqnarray}
From this equation, the nearest and next-nearest correlation
functions can be obtained readily as
$<\sigma_{i}\sigma_{i+1}>=\tanh(K)$ and
$<\sigma_{i-1}\sigma_{i+1}>=\tanh^{2}(K)$ respectively. It is
important to notice that the nearest neighbor correlation function
is the same as previously obtained result which is based on
configurational space equivalence of $\{\sigma_{i}\sigma_{i+1}\}$
and $\{s_{i}\}$, in which no boundary condition is used, here
$s_{i}=\pm1$. The nearest and next nearest neighbor correlation
relations are also the same as those of obtained from transfer
matrix treatment with periodic boundary condition.

Applying Block-spin transformation again on the partition function
in Eq.(8), one can easily obtain the following equivalent
partition function
\begin{eqnarray}
Q=\sum_{\{\sigma\}}f_{1}(K)^{N/2}f_{2}(K)^{N/4}e^{g_{2}(K)\sum
\sigma_{i-1}\sigma_{i+1}}
\end{eqnarray} Where,
\begin{eqnarray*}
&&g_{2}(K)=\frac{1}{2}\ln(\cosh(2g_{1}(K))),{}\nonumber
\\&&
f_{2}(K)=2[\cosh(2g_{1}(K)]^{1/2}.
\end{eqnarray*}
Taking logarithmic derivative of the both sides of equation
Eq.(12) with respect to $K$ leads to
\begin{eqnarray}
&&<\sigma_{i}\sigma_{i+1}>=\frac{\tanh(2K)}{2}+{}\nonumber
\\&&\frac{(\sinh^{2}(2K))\tanh(2K)}{6+2\cosh(4K)}<\sigma_{i-2}\sigma_{i+2}>
\end{eqnarray}
Using the relation in Eq.(10), the last relation can be written in
the following form
\begin{eqnarray}
<\sigma_{i-1}\sigma_{i+1}>=\frac{(\sinh^{2}(2K))}{3+\cosh(4K)}(1+<\sigma_{i-2}\sigma_{i+2}>).
\end{eqnarray}
It is not hard to see that the function on the right side of the
equation can easily be written equivalently as
\begin{eqnarray}
\frac{(\sinh^{2}(2K))}{3+\cosh(4K)}=\frac{\tanh^{2}(K)}{1+\tanh^{4}(K)}.
\end{eqnarray}
Substituting the equivalent form and rearranging the last equation
leads to
\begin{eqnarray}
\frac{<\sigma_{i-1}\sigma_{i+1}>}{1+<\sigma_{i-2}\sigma_{i+2}>}=\frac{\tanh^{2}(K)}{1+\tanh^{4}(K)}.
\end{eqnarray}
This last equation indicates that the fourth nearest neighbor
correlation function $<\sigma_{i-2}\sigma_{i+2}>$ is equal to
$(\tanh^{4}(K))$. It is also important to notice that
$<\sigma_{i-1}\sigma_{i+1}>=(\tanh^{2}(K))$ result is once more
recovered.

Since now, we have applied Block-spin transformation two times and
proved that $<\sigma_{i}\sigma_{i+k}>=(\tanh^{k}(K))$ for $k=1,2$,
and $4$. Intuitively, one can expects that applying Block-spin
transformation once more leads to a relation between fourth
nearest neighbor and eight nearest neighbor correlation functions.
The form of the expected relation might also be in the form of
Eq.(11) and Eq.(16). But doing so does not provide a general
proof. In addition, it inevitably creates more mathematical
complications. So it is apparently necessary to obtain a general
proof of the relations between correlation functions. We derived
very useful very relation to overcome the inevitable mathematical
complications presented in Appendices A. Using the obtained result
in the appendices A., which is expressed as
\begin{eqnarray}
\frac{<\sigma_{i}\sigma_{i+j}>}{1+<\sigma_{i}\sigma_{i+2j}>}=\frac{1}{2}\tanh(2g_{n}(K)),
\end{eqnarray}
one can obtain the following relation for $j=4$,
\begin{eqnarray}
\frac{<\negthinspace\negthinspace
\sigma_{i\negthinspace-\negthinspace2}\sigma_{i \negthinspace+
\negthinspace2}>}{\negthinspace1+\negthinspace<\negthinspace\sigma_{i-4}
\sigma_{i+\negthinspace4}>}=\negthinspace\frac{1}{2}\tanh(\ln(\cosh(\ln(\cosh(2K)\negthinspace)\negthinspace)\negthinspace)\negthinspace).
\end{eqnarray}
It is not hard to see the equivalence of the function on the right
hand side of the last equation  with
$\frac{\tanh^{4}(K)}{1+\tanh^{8}(K)}$. Thus,  one can obtain the
following following relation
\begin{eqnarray}
\frac{<\sigma_{i-2}\sigma_{i+2}>}{1+<\sigma_{i-4}\sigma_{i+4}>}=\frac{\tanh^{4}(K)}{1+\tanh^{8}(K)}.
\end{eqnarray}
Using the mathematical induction concept as we all resort so
often, this final result can be considered the proof of the
following general relation
\begin{eqnarray}
\frac{<\sigma_{i}\sigma_{i+j}>}{1+<\sigma_{i}\sigma_{i+2j}>}=\frac{\tanh^{j}(K)}{1+\tanh^{2j}(K)}.
\end{eqnarray}
From this last relation, the equation
$<\sigma_{i}\sigma_{i+j}>=\tanh^{j}(K)$ for $j=2^{n}$, here
$n=0,1,2,3,4,...$, can be suggested for the general case.

We think that this equations are valuable in that they are
obtained relevantly by the use of a different unique method other
than the transfer matrix method. In addition they are going to be
helpful to obtain relevantly the order parameter or the average
magnetization in the limit of as $\pm h$ goes to zero. At first
sight, it is amusing to take the limit from the zero field
correlation functions but it can be relevantly achieved in the
following section.
\section{\label{sec:level1}Obtaining The order parameter from zero field correlation function}
In this section we present and obtain a proper relation for the
order parameter. Indeed, the final result of this section have
already be used in many research \cite{Baxter4} without having any
proper discussion. In another words, the relation between
infinitely apart spins correlation and order parameter, $\lim_{N
\to \infty}<\sigma_{i}\sigma_{i+N}>=<\sigma>^{2}$, is believe to
be true, but not proven. Of course, representing a proof for this
relation in some sense is confusing and difficult without the use
of some physical intuition. Therefore our aim in this section not
to present the proof of this relation, but instead to give
plausible discussion about the relevancy of this relation. Now,
let us start to our discussion by writing the partition function
of 1D Ising chain in the presence of external field for nearest
neighbor interactions.
\begin{eqnarray}
Q=\sum_{\{\sigma\}}e^{K\sum\sigma_{i}\sigma_{i+1}+h\sum\sigma_{i}}
\end{eqnarray}
The order parameter and the correlation function can be defined
also as
\begin{eqnarray}
<\sigma_{i}>=\frac{1}{Q}\sum_{\{\sigma\}}\sigma_{i}e^{K\sum\sigma_{i}\sigma_{i+1}+h\sum\sigma_{i}}
\end{eqnarray}
\begin{eqnarray}
<\sigma_{i}\sigma_{i+j}>=\frac{1}{Q}\sum_{\{\sigma\}}\sigma_{i}\sigma_{i+j}e^{K\sum\sigma_{i}\sigma_{i+1}+h\sum\sigma_{i}}
\end{eqnarray}
Now consider the average magnetization relation in the $h=0$ case.
Due to spin filliping symmetry, the value of average magnetization
is expected to be zero in this case so this consideration is not
proper to define the order parameter in the limit of $\pm h$ goes
to zero. However, if the two spin correlation function relation is
considered the term $\sigma_{i}\sigma_{i+j}$ in the sum, breaks
the spin filliping symmetry a lot stronger than the term
$h\sum{\sigma_{i}}$ in the limit of $h$ goes to zero. Therefore,
we can assume intuitively that the zero field two point
correlation function relations are physically suitable to
calculate zero magnetic field order parameter. In an other words,
$<\sigma_{i}\sigma_{i+j}>|_{h=0}$ is not different from $\lim_{\pm
h \to 0} {<\sigma_{i}\sigma_{i+j}>}$. Now considering two very far
(or infinitely) apart spins, which are, of course, uncorrelated
from classical physic perspective, one can readily writes the
following equation,
\begin{eqnarray}
\lim_{j \to \infty}<\sigma_{i-j}\sigma_{i+j}>=\lim_{h \to
0}<\sigma_{i-j}><\sigma_{i+j}>.
\end{eqnarray}
Thus the order parameter or the average magnetization for per spin
as $h$ goes to zero can be defined as
\begin{eqnarray}
<\sigma_{i-j}>=(\tanh(K))^{j}.
\end{eqnarray}
and as $-h$ goes to zero
\begin{eqnarray}
<\sigma_{i-j}>=-(\tanh(K))^{j}.
\end{eqnarray}

It is important to notice that the average magnetization for per
spin is independent of the space position of the spin or the
lattice site indicating that $<\sigma_{i}>$ is equal to
$(\tanh(K))^{j}$ for any $i$ or any lattice site. Notice also that
the ideal uncorrelated two spin case can be satisfied only when
$j$ goes to infinity. This means that the only none zero value of
the order parameter is only passible for infinitely large values
of $K$. This result is totally in agrement with the result
obtained from transfer matrix calculation with periodic boundary
condition. In the language of phase transition theory, one can
also restated the last remark as second order phase transition is
only possible in the limit of $K$ goes to infinity. Indeed, in
this case, the value of the average magnetization is equal to $1$
as $h\rightarrow{0}$, and -1 as $-h\rightarrow{0}$.

In conclusion, we have studied $1D$ Ising chain in the absence of
external magnetic field. In the introduction section, we have
given some discussion about a possible relevant physical picture
to obtain average magnetization from a partition function in the
absence of external field. Several ambiguous points related to
taking properly the limit of order parameter as $h$ goes to zero
have been dealt with a heuristic manner only. We did not go into
the details of mathematical rigor since our aim was to show that
the relation $<\sigma_{i}>^{2}=\lim_{n \to \infty}
<\sigma_{i}\sigma_{i+n}>$ could be a relevant consideration to
obtain the average magnetization in the absence of the external
field. Furthermore, an alternative solution  with no boundary
condition for 1D Ising chain was obtained with the help of
successive Block-spin transformations. We think that the average
magnetization definition is quite acceptable. In other words,
infinitely long-range two spin correlations function  can be used
to obtain average magnetization. In addition, no boundary
condition have been used in our treatment. Since the results
obtained in this paper with successively applying Block-spin
transformation and the transfer matrix method are the same, we
might claim that the boundary conditions, either periodic or free,
have no effect on the 1D Ising chain. For future considerations,
we think that the developed method in this paper for $1D$ Ising
chain can be also applied to $2D$ or $3D$ Ising systems.
Apparently, its application to higher dimensional Ising system can
not be as direct as 1D case due to the complications arise from
four spin and six spin correlation function around central spin
which inevitable appears if Block-spin transformation applied to
higher dimensional situations. I also think that 2D and 3D Ising
systems can be treated more accurately and efficiently if the
method developed in this paper combine with the real space
renormalization group assumptions.

\subsection{Appendices}
In order to obtain Eq.(18), it is relevant to present the
partition function in a general case. The general case means here
that writing the partition function after applying the Block-spin
transformation $n$ times. So the partition function can be written
in the general case as
\begin{eqnarray*}
Q=\sum_{\{\sigma\}}f_{0}^{N}f_{1}^{N/2}...f_{n}^{N/(2^n)}e^{g_{n}(K)\sum
\sigma_{i}\sigma_{i+j}} .
\end{eqnarray*}
Where, $g_{0}(K)=K$, $f_{0}(K)=1$ and $j=2^{n}$. For our purpose,
we also need to write the partition function after applying
Block-spin transformation $(n+1)$ times, which is apparently equal
to
\begin{eqnarray*}
Q=\sum_{\{\sigma\}}f_{0}^{N/2}f_{1}^{N/4}...f_{n-1}^{N/(2^{(n+1)})}e^{g_{n+1}(K)\sum
\sigma_{i}\sigma_{i+j^{'}}} .
\end{eqnarray*}
where $j^{'}=2^{n+1}=2j$. Now calculating the logarithmic
derivative of both of these relations and doing some algebra among
them leads to
\begin{eqnarray*}
\frac{d\ln(f_{n+1})}{dK}+\frac{dg_{n+1}}{dK}<\negthinspace\negthinspace
\sigma_{i}\sigma_{i+j^{'}}\negthinspace\negthinspace>=\frac{1}{2}\negthinspace\frac{dg_{n}}{dK}
<\negthinspace\negthinspace\sigma_{i}\sigma_{i+j}\negthinspace\negthinspace>\negthinspace\negthinspace\negthinspace
\end{eqnarray*}
where $g_{n+1}(K)$ and $f_{n+1}(K)$ have the following relations
\begin{eqnarray*}
&&g_{n+1}(K)=\frac{1}{2}\ln[\cosh(2g_{n})],{}\nonumber
\\&&
f_{n+1}(K)=2[\cosh(2g_{n})]^{1/2}.
\end{eqnarray*}
Substituting these relations into the last equation leads to the
following equation
\begin{eqnarray*}
(1+<\sigma_{i}\sigma_{i+j^{'}}>)\frac{dg_{n+1}}{dK}=\frac{1}{2}<\sigma_{i}\sigma_{i+j}>\frac{dg_{n}}{dK}
\end{eqnarray*}
rearranging this equation leads to
\begin{eqnarray*}
\frac{<\sigma_{i}\sigma_{i+j}>}{1+<\sigma_{i}\sigma_{i+j^{'}}>}=\frac{1}{2}\frac{dg_{n+1}}{dg_{(n)}}
\end{eqnarray*}
Notice that the derivative $\frac{dg_{(n+1)}}{dg_{(n)}}$ is equal
to $\tanh(2g_{(n)})$. Thus the last equation can be expressed as
\begin{eqnarray}
\frac{<\sigma_{i}\sigma_{i+j}>}{1+<\sigma_{i}\sigma_{i+2j}>}=\frac{1}{2}\tanh(2g_{n}(K)),
\end{eqnarray}
where we have used $j^{'}=2j$. This final equation is very useful
to obtain the proof of the general correlation function relations
from the point of view of mathematical induction.

\end{document}